\documentclass{aastex}

\shorttitle{A CMB Map from BEAST}
\shortauthors{Meinhold et al.}

\begin{document}

\title{A Map of the Cosmic Microwave Background from the 
BEAST Experiment.}

\author{Peter R. Meinhold \altaffilmark{1,2},
Marco Bersanelli\altaffilmark{6},
Jeffrey Childers\altaffilmark{1},
Newton Figueiredo\altaffilmark{5}, 
Todd C. Gaier\altaffilmark{3}, 
Doron Halevi\altaffilmark{1},
Miikka Kangas\altaffilmark{1}, 
Charles R. Lawrence\altaffilmark{3},
Alan Levy\altaffilmark{1},
Philip M. Lubin\altaffilmark{1,2},
Marco Malaspina\altaffilmark{8},
Nazzareno Mandolesi\altaffilmark{8},
Joshua Marvil\altaffilmark{1},
Jorge Mej\'{\i}a\altaffilmark{4}, 
Paolo  Natoli\altaffilmark{7},
Ian O'Dwyer\altaffilmark{9},
Hugh O'Neill\altaffilmark{1},
Shane Parendo\altaffilmark{1},
Agenor Pina\altaffilmark{5},
Michael D. Seiffert\altaffilmark{3}, 
Nathan C. Stebor\altaffilmark{1},
Camilo Tello\altaffilmark{4},
Fabrizio Villa\altaffilmark{8},
Thyrso Villela\altaffilmark{4}, 
Lawrence A. Wade\altaffilmark{3,11}
Benjamin D. Wandelt\altaffilmark{10} ,
Brian Williams\altaffilmark{1},
Carlos Alexandre Wuensche\altaffilmark{4}
}

\altaffiltext{1}{Physics Department, University of California, Santa Barbara, CA 93106}
\altaffiltext{2}{University of California, White Mountain Research Station, CA 93514}
\altaffiltext{3}{Jet Propulsion Laboratory, California Institute of Technology, Oak Grove Drive, Pasadena, CA 91109}
\altaffiltext{4}{Instituto Nacional de Pesquisas Espaciais Divis\~ao de Astrof\'{\i}sica Caixa Postal 515,12210-070 -  S\~ao Jos\'e dos Campos, SP Brazil}
\altaffiltext{5}{Universidade Federal de Itajub\'a Departamento de F\'{\i}sica e Qu\'{\i}mica Caixa Postal 50 37500-903 - Itajub\'a, MG Brazil}
\altaffiltext{6}{University of Milano, via Celoria 16, 20133 Milano, Italy}
\altaffiltext{7}{Dipartimento di Fisica e sezione INFN, Universit\`a di Roma "Tor Vergata", Rome, Italy}
\altaffiltext{8}{IASF-CNR sezione di Bologna, via P.Gobetti, 101, 40129 Bologna, Italy}
\altaffiltext{9}{Department of Astronomy, University of Illinois at Urbana-Champaign Urbana, IL 61801}
\altaffiltext{10}{Department of Physics, University of Illinois at Urbana-Champaign Urbana, IL 61801}
\altaffiltext{11}{Department of Applied Physics, California Institute of Technology, Pasadena, CA   91125}

\begin{abstract}
We present the first sky maps from the BEAST (Background Emission 
Anisotropy Scanning Telescope) experiment. BEAST consists of a 2.2 meter off axis 
Gregorian telescope fed by a cryogenic millimeter wavelength focal 
plane currently consisting of 6 Q band (40 GHz) and 2 Ka band (30 
GHz) scalar feed horns feeding cryogenic HEMT amplifiers. 
Data were collected from two balloon-borne flights in 2000,
followed by a lengthy ground 
observing campaign from the 3.8 Km altitude University of California 
White Mountain Research Station. This paper reports the initial results 
from the ground based observations.
The instrument produced an annular map covering the sky from $33\degr \le \delta \le  42\degr$. 
The maps cover an area of 2470 square degrees with an effective resolution of 23 arcminutes FWHM at 40 
GHz and 30 arcminutes at 30 GHz. The map RMS (smoothed to 30 arcminutes and excluding galactic foregrounds) 
is $54\pm5 \mu$K at 40 GHz.  Comparison with the instrument noise gives a cosmic signal RMS contribution of $28\pm3 \mu$K. An estimate of the actual
CMB sky signal requires taking into account the $\ell$ space filter function of our experiment and
 analysis techniques, carried out in a companion paper (O'Dwyer et al. 2003).
In addition to the robust detection of CMB anisotropies, we find a strong correlation between small portions of our maps and features in recent H$\alpha$ maps (Finkbeiner 2003). 
In this work we describe the data set and analysis techniques leading to the maps, including data 
selection, filtering, pointing reconstruction, mapmaking algorithms and systematic effects. A detailed 
description of the experiment appears in Childers et al. (2003).
\end{abstract}


\section{Introduction}
The Cosmic Microwave Background (CMB), interpreted as relic radiation from an early hot dense phase of the
universe, is generally acknowledged as one of our primary probes of conditions in the early universe(Hu and Dodelson 2002 and references therein).
Mapping the temperature anisotropies of the CMB has been the subject of intensive effort over the past two decades, with the
results of many studies published  within the last 3 years:  BOOMERANG (Netterfield et al. 2001), 
MAXIMA(Lee et al. 2001), CBI (Mason et al 2002), DASI(Halverson et al. 2002), Archeops (Benoit et al. 2002), 
and ACBAR(Kuo et al. 2002) among others (see Bersanelli et al 2002 for a recent review).

Temperature anisotropy measurements have been used to calibrate
the amplitude of the initial density perturbations and constrain 
the primordial spectral index (Gorski et al. 1996), measure the 
overall density of the universe(De Bernardis et al. 2001, etc.), 
and recently to make detailed comparisons between competing scenarios 
for structure formation, leaving a rather narrow range of consistent 
adiabatic inflationary models.  In combination with recent constraints 
from supernova SNIa and Large Scale Structure (LSS) data, CMB temperature 
anisotropy data strongly suggest models with a large fraction of the total 
energy density in a cosmological constant or other 'dark sector'.

CMB data are normally reduced to an estimate of the angular power spectrum of fluctuations for comparison to theoretical predictions.
The salient features of the currently allowed adiabatic inflationary models are a series of  'acoustic' peaks and troughs in the power spectrum
from $\ell$ (spherical harmonic multipole) of about 200  and up.(Bond \& Efstathiou 1987)  The positions and amplitudes of these peaks pin down the 
details of the cosmological models.  Each of the aforementioned experiments was  designed to target a particular region of '$\ell$-space' or portion of the angular power spectrum.  Sensitivity to high $\ell$ modes is limited by the angular resolution of the telescope.
Resolution in $\ell$-space is limited by the size of the region studied. Amplitude uncertainties at all $\ell$ are limited by the detector noise and
integration time. At low $\ell$ values in particular, sample variance can increase the amplitude uncertainty of a measurement, an effect which
can only be reduced by increasing sky coverage. BEAST, in particular the observing campaign described in this paper, has been optimized to study the first of these peaks in detail.

\subsection{Instrument Overview}

BEAST is a mixed Q-band (38-45 GHz) and Ka-band (25-35 GHz) focal plane 
that views the sky through a 2.2 m off-axis telescope followed by a modulating flat mirror.  Primarily funded by NASA, BEAST is largest aperture CMB telescope flown to date. 
The instrument is described in detail in 
Childers et al. (2003), while the optical design is described in Figueiredo et al. (2003).
BEAST is most similar to one of our previous balloon 
borne telescopes, HACME (Staren et al. 2000; Tegmark et al. 2000) but with a much larger 
telescope and a focal plane array. 
After two balloon flights in 2000, BEAST was  reconfigured to take advantage of the UC White 
Mountain Research Station (WMRS) site on Barcroft mountain, CA (altitude 3800 m ).
The data 
described here were taken with 6 elements of the array operational, 2 Ka-band and 4 Q-band detectors. 2 of 
the Q-band detectors are installed on an Ortho-Mode Transducer (OMT) using a single horn split into 
vertical and horizontal polarizations. The other horns have single polarization receivers, oriented
at various angles. The telescope was installed 
in an unused building, modified so the roof can be rolled off. BEAST is aligned at 74 degrees East of 
North, and can view nominal central elevation angles from 60 to greater than 90 degrees.
The instrument was fully installed and operational at Barcroft in July, 2001, and took data nearly 
continuously until December 2001. Two more weeks of data were obtained in February 2002 and a longer campaign took place from 
August through October of 2002.

\subsection{Observing strategy}

BEAST views the sky through a large (2.5 meter diameter) spinning flat mirror, tilted 
2.2 degrees from its rotation axis.  As the flat spins this tilt moves the field of view of the telescope around a 
nearly elliptical path on the sky of diameter about 9 degrees. 
The 5 operational feeds thus follow the path shown in Figure 1
for the normal telescope elevation angle of 90.4 degrees. Two main strategies were employed in taking data. 
The first involved scanning the nominal (central) elevation of the telescope from 80 to 90 degrees and back 
continuously, the second method was to keep the elevation angle fixed near 90 degrees and allow Earth 
rotation to provide the mapping scan. All of the data reported here are from the 'fixed elevation' data set. 
This choice was made primarily for the simpler and more robust pointing solution obtained in fixed 
elevation mode, and the fact that there were no significant improvements in systematic tests gained from 
the elevation scans. The resulting map is an annulus around the North Celestial Pole (NCP) of width $\sim 10$ degrees, around
declination 37 degrees. Figure 2 shows the entire Q-band map with the beam trajectories overlaid.

The output of each radiometer is amplified, ac coupled with a highpass filter time constant of 15 seconds, and then 
integrated in an 'ideal' or boxcar integrator with nearly 100\% duty cycle. The integration time is set for 
$ 800 \mu$s and digitized to 16 bits.

The angular position of the flat mirror (rotating at 2 Hz) is measured simultaneously so the data 
can be synchronized with the instantaneous position of the relevant telescope beam. The first step of analysis
involves rebinning the raw 1250 Hz data stream into 'optical sectors', corresponding to 250 angular positions of 
the flat mirror. Each 'optical sector' therefore, identifies an angular position of the flat, which is related by the telescope 
geometry to a precise azimuth-elevation pointing for each of the 5 active detector elements.  These pointing data 
are converted to celestial coordinates and then ultimately to HEALPix  (Hierarchical Equal Area isoLatitude Pixelisation, Gorski et al. (1998)) pixel numbers. The resulting optical sector 
data stream has an effective sampling rate of approximately 450 Hz.
The raw data amount to approximately 3.7 Gigabytes per day.

\subsection{Calibration}
The receivers are calibrated hourly by passing an ambient temperature microwave absorber between the 
feeds and the secondary mirror. The output of the acquisition system is then corrected for the effects of the 
ac coupling in addition to a small (order 10\%) correction for compression of the detector outputs while 
viewing the ambient target. The difference between the load 
and the sky temperatures is then used to compute the calibration coefficients, measured in K/V. 
This hourly calibration constant is applied to the data during pipeline analysis.  Calibration and effective 
beamsize were confirmed independently using Cygnus A as the celestial source.  Overall calibration uncertainty 
is $\pm$10\%. Calibration and beamsize issues are covered more completely in Childers et al. (2003).

\section{Data Set}

\subsection{Raw Data Statistics}

Figure 3 shows a sample section of time ordered data for a typical channel.  The top strip of small points is the raw 
data stream, including electrical and any radiometric offset. The large nearly sinusoidal component is due to the varying 
atmospheric contribution as the flat sends the beam through different elevations and therefore 
different optical depths.  This variation is approximately 60 mK peak-to-peak for this Q-band (38-45 GHz) channel.  
We obtain a high signal to noise estimate of the nearly constant atmospheric contribution for each 
revolution by binning all the data for each hour by optical sector to produce a 'template' signal for that hour.
We then subtract this template from every revolution in that hour to remove the first order atmospheric contribution. 
Note that systematic errors fixed in the telescope frame or phase locked to the flat revolution would be removed as well. Examples
of such effects are fixed sidelobe contamination and electrical noise synchronous with the flat motion. Since our
strategy keeps the telescope fixed, in principle any sidelobe illumination of the observatory or ground would be removed.
There is no direct evidence for any contribution except atmosphere.
The solid line plotted through the raw data points is the template for this hour, while the central strip of data points 
clustered near zero are the residuals from the template subtraction.  The lower strip of data points show the result
of highpass filtering the template-removed data set at 10 Hz
corresponding to about 6 degrees on the sky (effectively giving us a low $\ell$-space cutoff).
These data have been offset from zero for clarity.
The combination of 1 hour template removal and steep 10 Hz highpass filter
is a reasonable compromise between 
between extending the low $\ell$ coverage 
and minmizing the increasing effects of 1/f instrument and atmospheric noise contributions. Other filters have also 
been tried and will be reported on in the future.
Without the template removal, harmonics of the spin rate remain in the data and
can show up as striping in the maps. Hourly sections of data are our natural units since calibrations happen on the
hour, breaking the continuity of the data stream.  There appears to be no benefit to using longer sections for calculating the 
template: we have not yet investigated in detail the advantages of using shorter time periods for template subtraction.
Both hourly template removal and 10 Hz highpass filter were used in the analysis described below.

Estimates for the noise amplitude spectral density for this hour are given in Figure 4, calibrated
in $\mu\rm K\sqrt{\rm s}$. The large signal near 2 Hz and its higher frequency harmonics are 
due to the atmospheric contribution discussed above.  Notice that due to fact that the beam trajectory is not circular, 
and passes beyond zenith, the resulting atmospheric signal is distorted from sinusoidal, producing significant harmonics
of the spin frequency. Also shown on the plot is 
the noise density after template removal and 10 Hz highpass filtering.
The plot demonstrates the BEAST detector noise has
significant contributions from correlated, long memory (or ``1/f'') noise. The knee frequency of 
``1/f'' noise is defined as the frequency where the total
noise power spectral density (1/f + white) exceeds by a factor of two its white noise level. The latter
determines our sensitivity per unit time for angular scales near the beam size.  Table 1 
summarizes the receiver and atmospheric noise and signal performance for a typical 
accepted hour of data from this campaign. 

\begin{center}
\begin{table}[t]
\caption{Summary statistics for a typical hour of sky data~\label{noise_table}}
%\begin{tabular}{|c|c|c|c|c|}
\begin{tabular}{crrrr}

\tableline\tableline
Channel & Frequency & White Noise & $1/f$ Knee & Template p-p \\
 & (GHz) & ($\mu\rm K\sqrt{\rm s}$) & (Hz) & (mK) \\
\tableline
2& 38-45  & 697 & 38 & 84 \\
3& 38-45  & 1324 & 40 & 149 \\
4& 38-45  & 770  & 60 & 102 \\ 
6& 25-35  & 1106 & 105 & 76 \\
7& 25-35  & 966  & 110 & 107 \\
8& 38-45  & 642  & 58 & 87 \\
\tableline
\end{tabular}
\end{table}
\end{center}

\subsection{Pointing Reconstruction}
The BEAST scan strategy produces a large area map (2470 square degrees) with high resolution, and we require a precise 
pointing solution to associate every optical sector for every channel with a position in celestial coordinates. 
We know the geometry of the telescope, but not well enough to calculate the beam positions to sufficient 
accuracy for map reconstruction. Deviation from ideality of the optics
gives uncertanties on the "a priori" knowledge of the beam positions.
BEAST was placed in an existing 
building that did not allow us to move the telescope appreciably in azimuth, and we are unable to target the moon, sun 
or any planets for pointing verification. 
We bootstrapped the pointing model by starting with the known geometry, then flying a small airplane 
equipped with GPS over the system in a grid pattern. The spikes in the BEAST data due to the
airplane thermal emission were then 
correlated with the measured relative position of the airplane to make a crude improvement to the pointing model. 
Cygnus A as well as several other well known radio sources are included in our map, and we have 
been able to use these sources to further refine the pointing model, beamsize and calibration. Figure 5 is 
a portion of the final map (all Q-band channels averaged) showing Cygnus A and the Cygnus Loop region. 
The reconstruction returns an effective resolution of 23 arcminutes FWHM in Q-band.  This is in contrast 
with the previously measured telescope resolution of 19 arcminutes.  The smearing from the design 
resolution to the measured effective FWHM of 23 arcminutes is due to a combination of unmeasured
pointing errors (telescope flex, long term telescope sag),  residual errors in the pointing reconstruction
algorithm, smearing due to the finite HEALPix resolution of 6.9 arcminutes, and smearing from the inital flat rotation
sectors (about 6.7 arcminutes).  The pointing model is converted to a lookup table of azimuth and elevation for each horn,
sector and telescope elevation for use in the pipeline processing.

\subsection{Data Reduction and Analysis}
The first data reduction step was to reformat the data into optical sectors, and 
incorporate relevant pointing information into a single file (hereafter called 'level 0 FITS file'),
which conforms to the FITSIO standard. A single file was produced 
for each hour, including all channels, as well as telescope elevation, and other pointing information. In 
addition to rebinning to optical sectors we measured some statistics on these hourly sets to allow simple 
data culling later. In particular we measured the RMS of each channel for the hour as well as the RMS after 
highpass filtering the data at 10 Hz. Transients (generally due to thermal emission from airplanes crossing the instruments field of view) were 
removed and replaced with adjacent data. 
These points were flagged as 'unusable' for later analysis stages.
Level 0 processing was:1) Import 1 hour Receiver and Pointing data, 2)  Calculate statistics (RMS and highpass filtered RMS),
3) Rebin raw (not highpass filtered) data to Optical sectors, 4) Remove transients, flag bad or replaced points
5) Store data to FITS, save summary statistics in header.

The next step was to cull the data for bad weather and out of specification performance (ie when 
systems were off or under test or during calibrations). 
This was done by looking at the statistics of the highpass filtered data RMS in the level 
0 headers. Generally all the channels showed the same sensitivity to bad weather, and there was a clear knee in the distribution of RMS's for making the cut. 

%XX out of XX possible hourly files made the weather 
%cut. It should be noted that number of 'good ' hours per month or per total hours is not a reliable measure of 
%the conditions at WMRS generally. Many clear weather days were lost due to equipment problems or 
%scheduled shut downs, and during very bad weather, no data were taken at all. 

We then moved to 'level 1' data files, which integrated the pointing data into map pixel indices.
We chose to use the HEALPix (Gorski et al. 1998) set of standards and routines, partly for the inherent advantages of the pixelization 
scheme and partly for the extensive infrastructure of tools available from the HEALPix team for manipulating and
visualizing the resulting maps. This standardization and set of tools has been invaluable in performing all of our subsequent analysis \footnote {The HEALPix homepage is located at http://www.eso.org/science/healpix/.}.
Every sample for each channel was then associated with an azimuth-elevation pointing via the lookup table mentioned 
above, then to a celestial coordinate pointing using the known observation time and location, thence to a HEALPix index 
using the HEALPix IDL library routines. We chose to use Nside=512, with a resolution corresponding to 6.9 arcminute 
pixels as a reasonable oversampling of our beamsize.  Level 1 data files then included time, 6 columns of data, 1 column 
of  'goodness-of-data' flag, and 6 columns of HEALPix indices.

To produce a map, we 1)  Read in Level 1 file, 2) Calculate and subtract hourly atmospheric template from all channels, 
3) highpass filter all channels at 10 Hz, 4) Bin into daily map (ignoring flagged points, saving statistics), 5) Iterate for all 
hours to complete daily map for all channels, 6) Save daily map to disk, start next day
This generates a map with statistics for each day for each channel. We then co-added (averaged by pixel)
the daily channel maps into total data set channel maps, weighting by the average noise per pixel per day.
The Q-band channels were then co-added with appropriate noise weighting to create the total Q-band data maps, as
were the Ka channels.  In addition to these averaged maps, we made 'difference' maps, where we binned 1/2 of the days
into one map, the other 1/2 of the days into another. Then we generated the difference of these two maps on a pixel by pixel 
basis, dividing by 2 to maintain the noise statistics of the average maps. These difference maps give us a measure of the
noise in the maps vs the signal + noise in the average maps.  Some of these difference maps are displayed below.

\subsection{Maps and Statistics}
Figures 6 and 7 display two representative sections of the BEAST Q-band average map.  Each Figure contains 3 panels.
The top panel is the full
Nside=512 resolution map in gnomonic projection around $\delta$ = 38 degrees and $\alpha$=  0 and 160 degrees.
The second panel is this same map smoothed with a 0.5 degree FWHM gaussian.  The third panel is the 0.5 degree FWHM
smoothed difference map as a visual guide to the effective noise level in the smoothed average map.
\subsection{Noise properties}
Figure 8 shows the distribution of pixel values for the Q-band map, compared with a gaussian distribution. The 
distributions with and without the galaxy and point source cuts are shown. One measure of the noise and signal in the map is to calculate the standard deviation of the map (after removing foreground contaminated regions) and compare with the standard deviation in the difference map, which should have the same noise characteristics, but  no CMB signal.
For Q-band we find $254\pm 25 \mu \rm K$ at 6.9' pixels for the sum map and $252\pm 25 \mu \rm K$ for the difference map. Smoothing
to 30 arcminutes FWHM we get $54\pm 5 \mu \rm K$ for the sum and $47\pm 5 \mu \rm K$ for the difference, 
consistent with a signal RMS contribution of $26 \mu$ K (Rayleigh-Jeans units). Converting to blackbody spectrum and correcting for 3\% atmospheric
attenuation we obtain $28\pm3 \mu \rm K$ (quoted error dominted by calibration uncertainty). Note that this includes the map smoothing, 
and the complex BEAST filter produced by our AC coupling, beam shape, highpass filter, and 
template removal, so is not our best estimate of sky signal, but only an indicator of a statistically significant sky signal.

\subsection{Foregrounds}
Any sensitve CMB map of this size will cross regions of significant foreground contamination as is 
seen in our maps with the galactic crossings. Other than a few regions 
the map appears to be remarkably clean. By comparison with foreground maps, we have been able to determine 
a very conservative cut in galactic latitude which removes diffuse foregrounds, while for point sources we 
have used a strict point source finding algorithm operating on the data maps. Point sources were removed 
by excising a 1 degree diameter circle of pixels centered on each source.  The most interesting feature of 
our foreground contamination is a large structure near $\alpha$=60 degrees, $\delta$=36.5 degrees. The feature 
is 0.75 mK in the smoothed Q-band map and 1.4 mK in Ka-band, implying a power law spectral index of -1.9, 
consistent with free-free emission. We see a large correlation with a feature in the combined IRAS/DIRBE map 
(Finkbeiner et al. ), but no correlation at all with the Haslam map at 408 MHz.  Initially we thought this 
might be evidence for spinning dust grains, but a comparison with integrated H$\alpha$ emission showed a very 
large correlation, consistent with free-free emission. This comparison was made possible by the availability of 
a map of integrated H$\alpha$ emission produced by D. Finkbeiner (Finkbeiner, 2003) using data from WHAM (Reynolds 
et al. 2002), VTSS (Dennison et al. 1998), and SHASSA (Gaustad et al. 2001). 
The expected free-free emission is estimated by scaling from Bennett et al. (1992) which predicts $3 \mu \rm K$ 
/Rayleigh at 40 GHz. For our structure we measure $5.7 \mu$K/Rayleigh.  Since the correlations with the H$\alpha$ map 
were so much more evident than with Haslam and IRAS that we used the H$\alpha$ data to estimate our required galactic 
latitude cut: by cutting out data with $\mid b\mid$ less than $25 \degr$ we expect less than $10 \mu$K contamination 
from free-free emission.  This is a very 
conservative cut and we expect to refine this further to increase the accepted sky fraction.
Our resulting foreground removal template is shown in figure 9, shaded regions are used in the power spectrum analysis.

\section{Discussion}
We have produced extended ($\sim 2500$ square degree) sky maps with good angular resolution and control of systematics. 
Visual inspection of the smoothed maps shows unmistakable structure. Quantitative results for the power 
spectrum of CMB fluctuations which is consistent with our data are presented in a companion paper (O'Dwyer et al. 2003). 
In addition to cosmic structures we have interesting correlations with known astrophysical foregrounds. Further 
analysis will yield important information about the amplitude, spectra and possibly polarization of these foregrounds.

As an aid in interpreting our maps, and as a guide to our $\ell$-space filter, we have made a single realization of a 
'white' power spectrum sky (equal power on all scales), and then simulated the BEAST pipeline over that map. We created a
 full sky map with normally distributed noise in each pixel.  This map was then smoothed with a gaussian of FWHM=23 
arcminutes to simulate our effective beam.  Next we simulated the time ordered data expected from this map without 
additional noise by sampling with the BEAST pointing timestream. We ran the resulting timeline through the BEAST 
analysis pipeline, including template removal, filtering and binning. Finally we cut using the same foreground
template used in the actual data map. 'PseudoCl' power spectra (that is power spectra calculated assuming full 
sky coverage but actually covering only a particular region) were calculated for the full unsmoothed white input 
map, the smoothed white map, the smoothed and foreground cut white map, and the full  BEAST simulated smoothed 
and foreground cut map. 
The results are shown in figure 10. The two all sky power spectra have been scaled by the 
sky fraction (2.85\% of full sky) of the foreground cut template in order to put them onscale, and to demonstrate 
that the primary effect of limited sky coverage on our $\ell$-space filter is just amplitude. All of the smoothed 
maps display the high $\ell$ cutoff of our beam. Below $\ell$ of about 150, the input map power spectrum is 
virtually the same as the smoothed and cut sky power spectra, while the BEAST simulated map power spectrum 
displays the low $\ell$ cutoff of the combined template removal and 10 Hz highpass filter. The loss of sensitivity from 
$ \ell \approx 50$ to $\ell \approx 150$ is dominated by the highpass filter cutoff. Note that this analysis 
neglects correlations among different $\ell$s. Figure 11 gives an approximation to the $\ell$ space window 
function based on these calculation.

Power spectra from the data and cosmological parameter estimates will 
be presented in a companion paper (O'Dwyer et al. 2003).  

\section{Future Observations}
We have shown the utility of making large area 
sky maps with a simple scanning strategy and expect to add 
significantly more data in the near future. The data presented so far are 
the result of 27 days of "good data". Our experience so far is we can 
achieve "good data" more than 50\% of the days at our observatory. The 
site has proven to be extremely good in our wavelength range and the 
atmosphere stable. A paper discussing the site is in preparation.
One year of observations with the existing 
system only improved by new HEMT's should yield 7 $\mu$K errors per 
30 arcminute pixel on 10\% of the sky. The BEAST optics are 
precise enough to observe to nearly 1 THz so a variety of sky surveys 
are possible. At 90 GHz the beam size would be about 8.5 arc minutes 
FWHM while at 150 GHz it is 5 arcminutes.

\section{Acknowledgments}
It is a pleasure to thank K.  Gorski, D. Maino, D. Finkbeiner, S. Myers, T. Montroy and J. R. Bond for useful discussions. 
M. Lim and J. Staren made significant contributions early in the development of BEAST.
We wish to thank the White Mountain Rearch Station, particularly D. Trydahl, R. Masters, and M. Morrison,
for heroic efforts in getting BEAST sited. WMRS director F. Powell's enthusiastic support has
been critical to our success.  
BEAST was built with help from numerous talented UCSB physics undergraduates, as well
as the superb UCSB machine shop, and administrative staff. We wish to thank
E.Mattaini and E.Santambrogio from IASF Milano for there contributions.
We also want to thank the National Scientific 
Balloon Facility staff for their support on our two balloon campaigns leading up to the current ground based work.
The develpment and operations of BEAST were supported by NASA Office of Space Sciences, the National Science Foundation, 
University of California White Mountain Research Station, and the California Space Institute (CalSpace).
Production of the superb BEAST optics were made possible by personal and corporate support from
K. Kedward (UCSB ME), M. Pryor (COI), S. Dummer (Surface Optics), J. Wafer and T. Ives(Thin Film Technology),TRW, 
J. Anthony (Union Carbide) and Able Engineering. 
We had support from NRAO and M. Pospieszalski, as well as TRW in developing our HEMT amplifiers.
N.F. and A.P. were partially supported by CNPq grant \# 470531/2001-0. JM is supported by FAPESP
grants 01/13235-9 and 02/04871-1. 
TV and CAW were partially supported by FAPESP grant \# 00/06770-2.
TV was partially supported by CNPq grants \# 466184/00-0 and
302266/88-7-FA. CAW was partially supported by CNPq
grant \# 300409/97-4 and FAPESP grant \# 96/06501-4.
The research described in this paper was performed in part at the
Jet Propulsion Laboratory, California Institute of Technology, under a
contract with the National Aeronautics and Space Administration.

%refs

\clearpage
\begin{figure}
\plotone{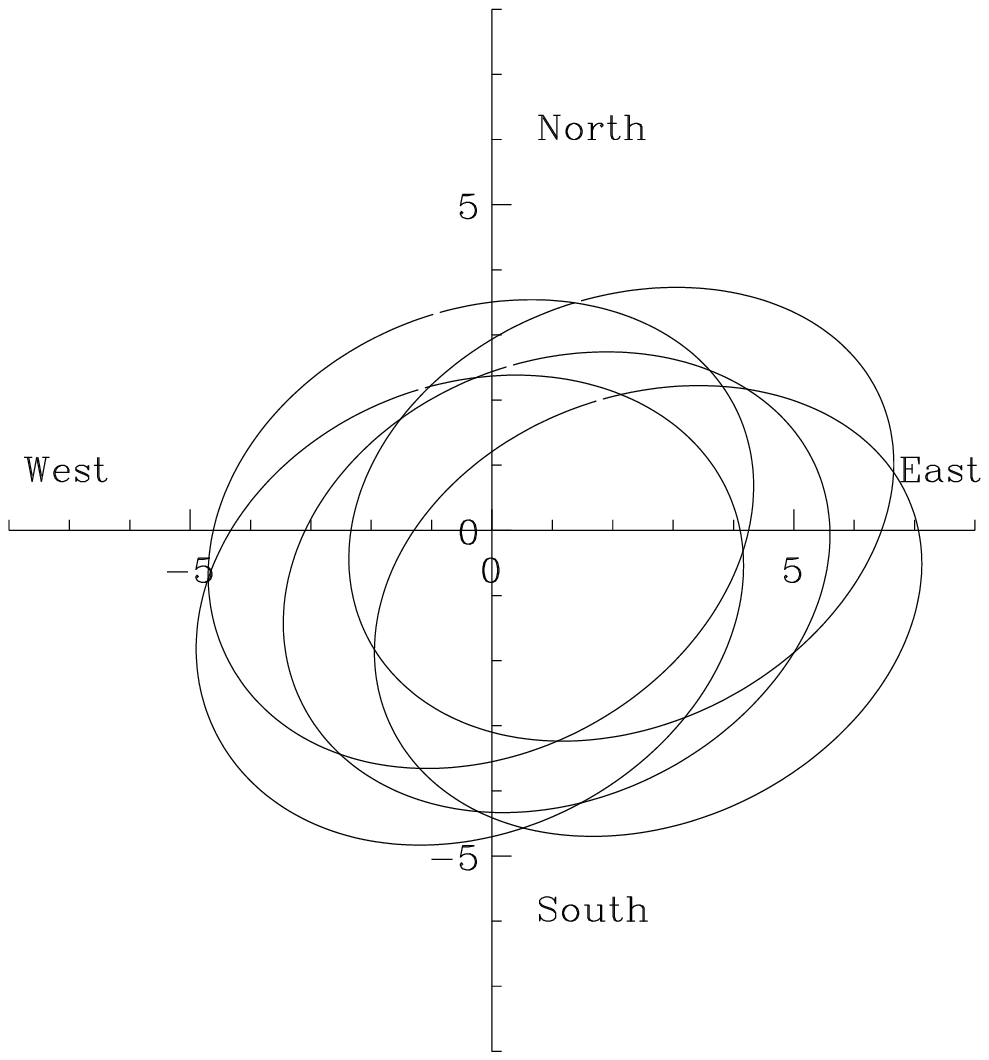}
\caption{Trajectories of the BEAST main beams. The figure is a polar 
plot about the local zenith. We subdivide the path of each beam
into 250 optical 'sectors' parameterized by the phase angle of 
the flat mirror which spins at about 2 Hz.\label{fig1}}
\end{figure}
\clearpage

\clearpage
\begin{figure}
\plotone{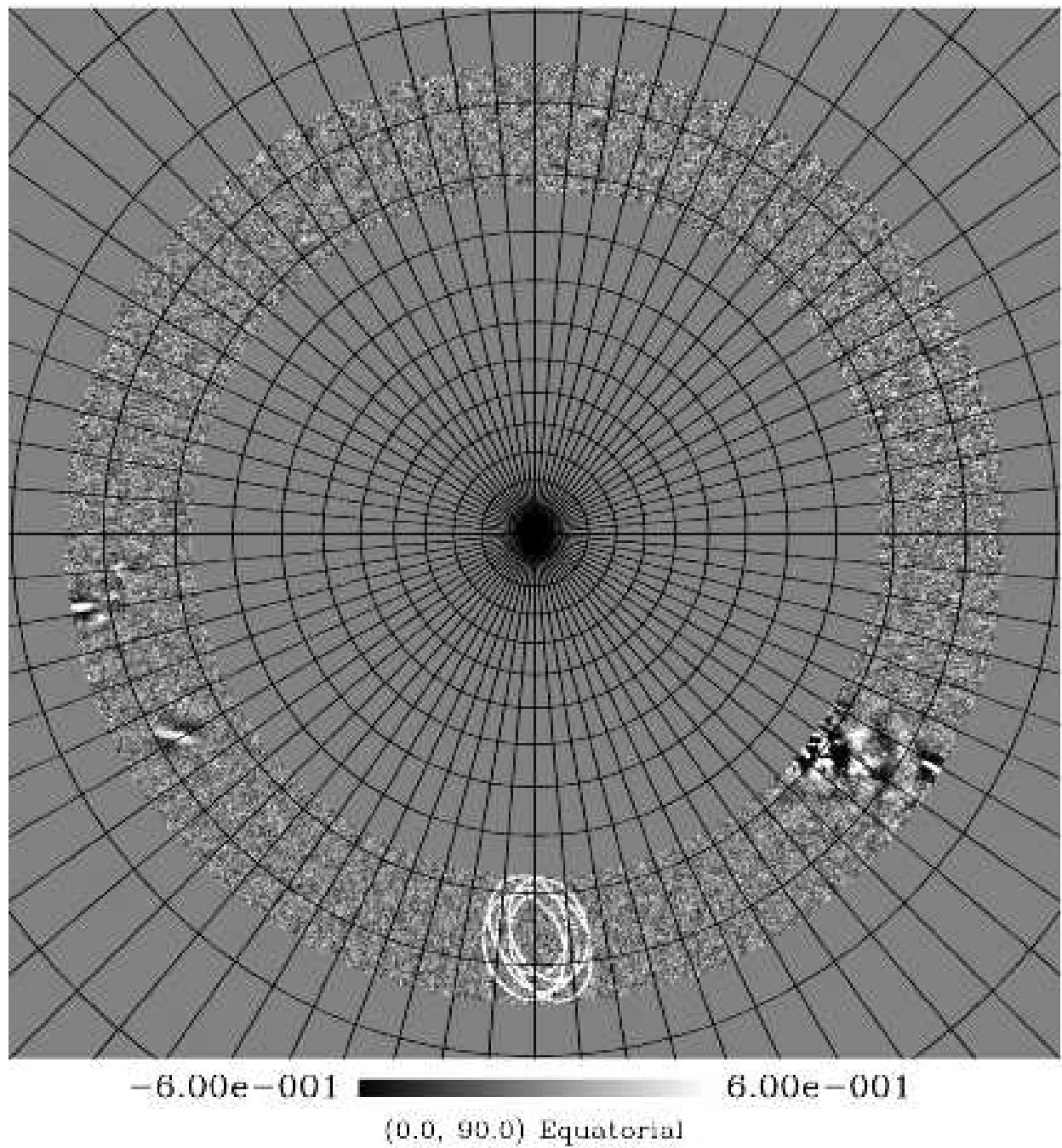}
\caption{The BEAST combined Q-band map. The figure is a 
gnomonic projection about the north celestial pole. The 
grid lines are 5 degrees in declination and 5 degrees in
right ascension. The trajectories of the 5 pixels during
1 revolution of the flat mirror are overlaid on the figure
near $\alpha$=0$\degr$ to illustrate the observing strategy.
The flat spins at about 2 Hz: the map includes data from over
4 million flat revolutions. Two galactic plane crossings are 
clearly visible on the plot, the brigher around $\alpha$=20$\degr$. 
Also visible is the large structure at $\alpha \approx 60 \degr$ 
found to be due to free free emission (see the text).\label{fig2}}
\end{figure}
\clearpage 

\clearpage
\begin{figure}
\plotone{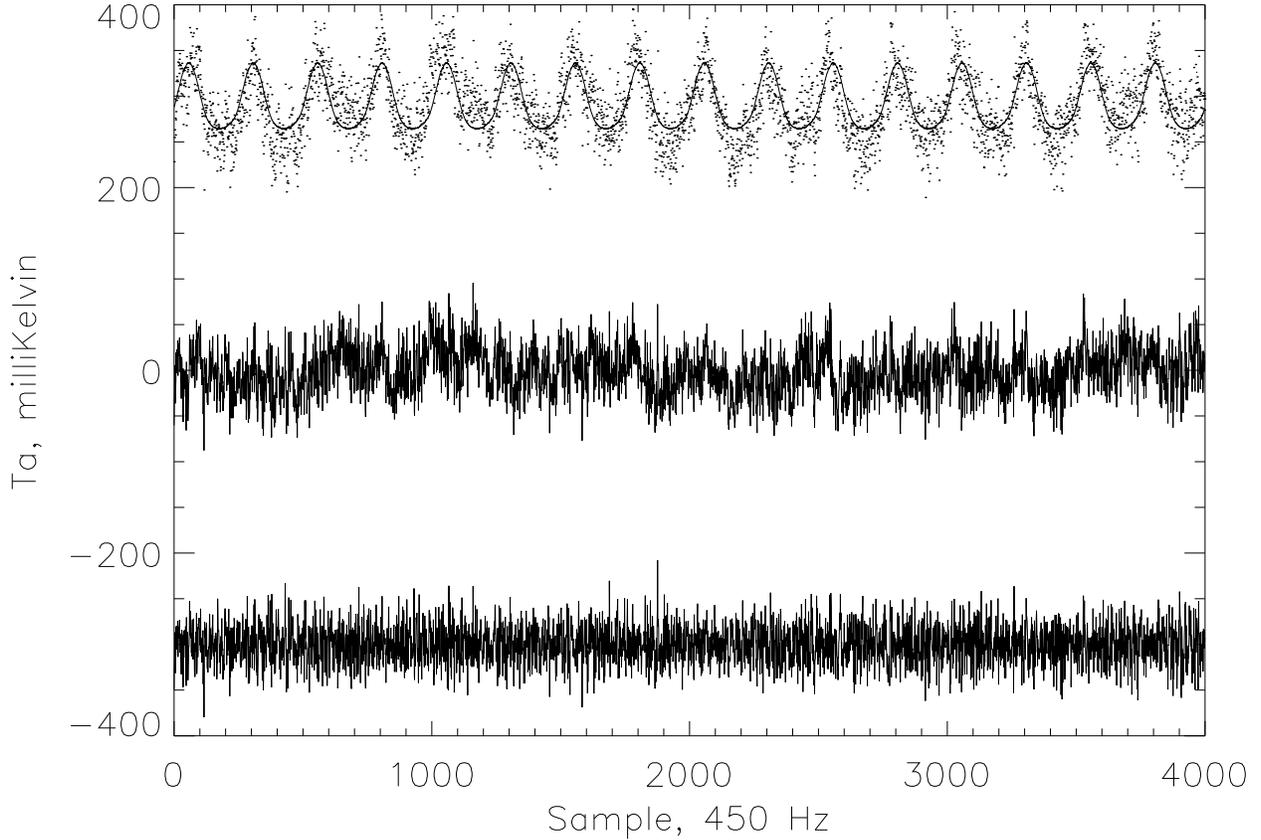}
\caption{Sample of time ordered data ($\approx 8$s). The top strip of small points is the raw 
data stream. The solid line through the points is the flat synchronous atmospheric
template obtained by averaging over this hour (see the text). The central strip of data points 
clustered near zero are the residuals from the template subtraction.  The lower strip 
of data points show the result of highpass filtering the template-removed data set at 10 Hz, 
offset from zero for clarity.\label{fig3}}
\end{figure}
\clearpage 

\clearpage
\begin{figure}
\plotone{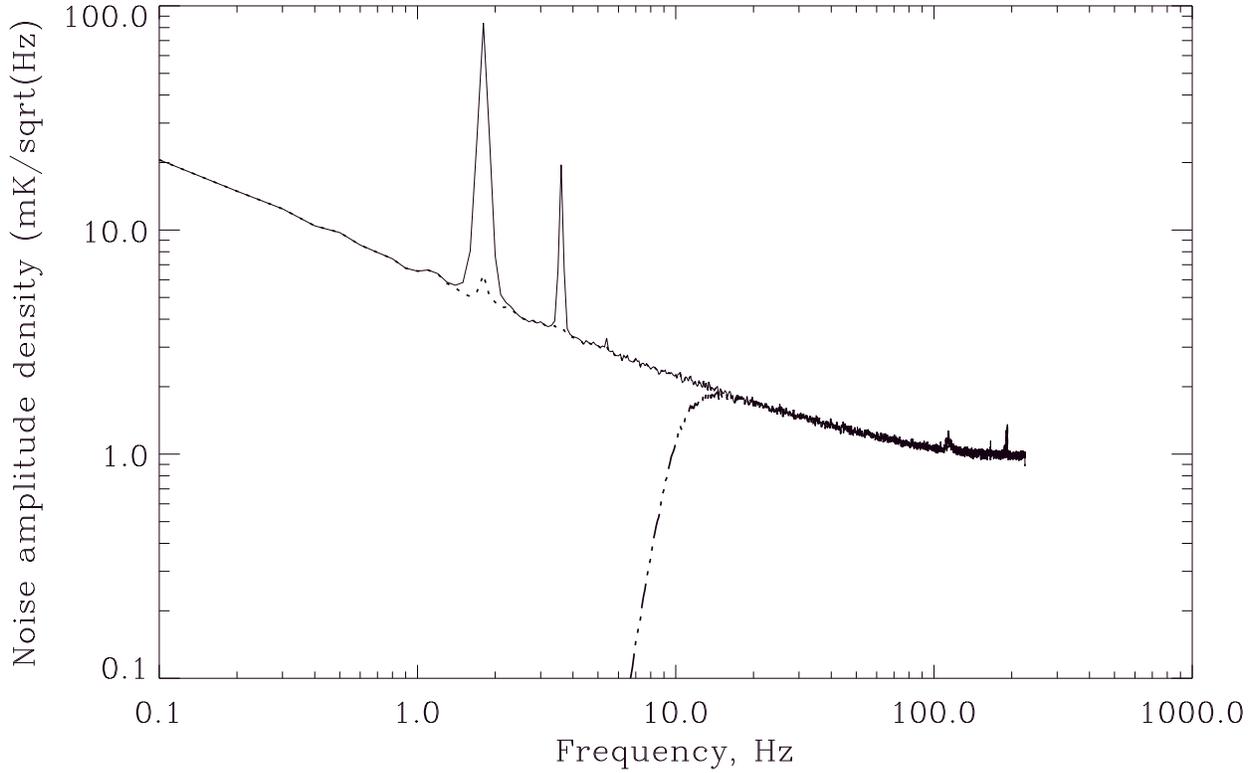}
\caption{Noise amplitude density spectrum from the sample hour of data. The white noise limit
at high frequency, the large 1/f component, the atmospheric signal at the spin rate and its 
harmonics, and excess noise at 120 Hz are all evident.  The noise spectrum for the template 
subtracted data (dotted line), and the spectrum after 10 Hz high are plotted as well(dashed line). \label{fig4}}
\end{figure}
\clearpage 

\clearpage
\begin{figure}
\plotone{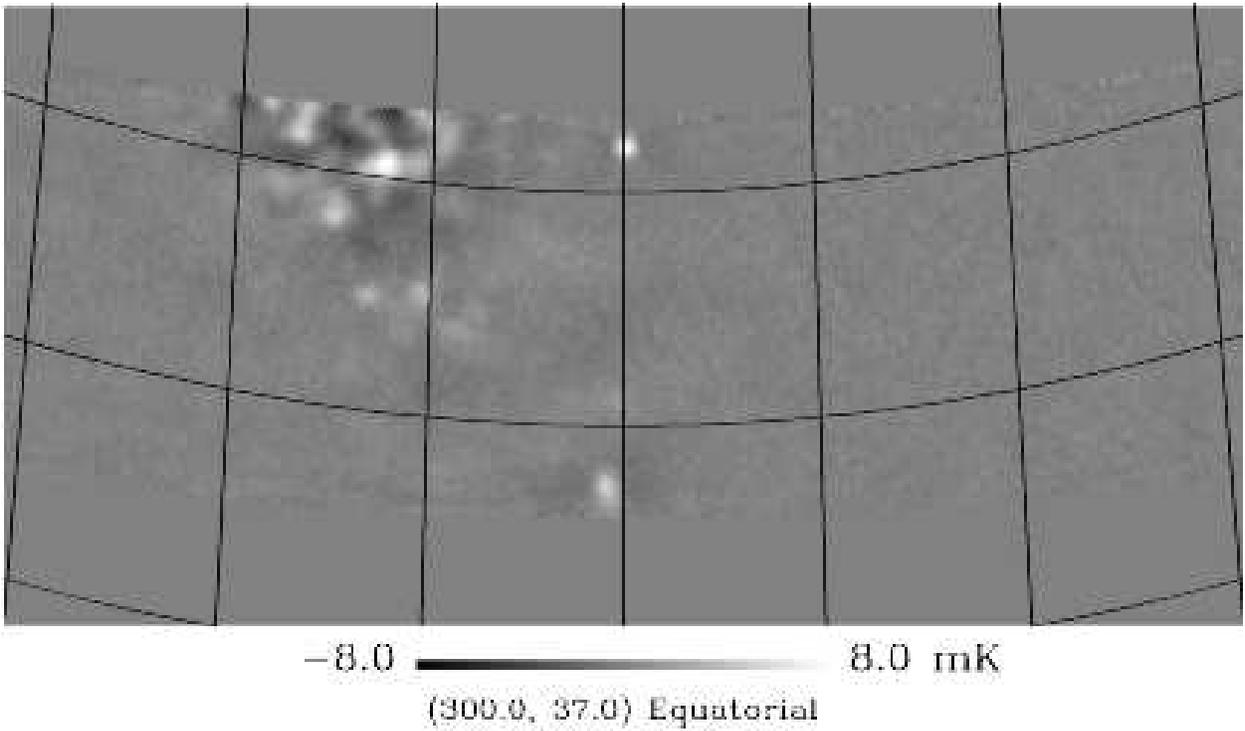}
\caption{ Gnomonic projection of the Q-band map near Cygnus A and 
the Cygnus Loop region. The grid spacing is 5 degrees in RA and Dec.
Cygnus A has been used as a pointing calibration as well as a cross
check of the effective beamsize and calibration
constant.  All of the sources seen here have been identified with known objects.\label{fig5}}
\end{figure}
\clearpage 

\clearpage
\begin{figure}
\plotone{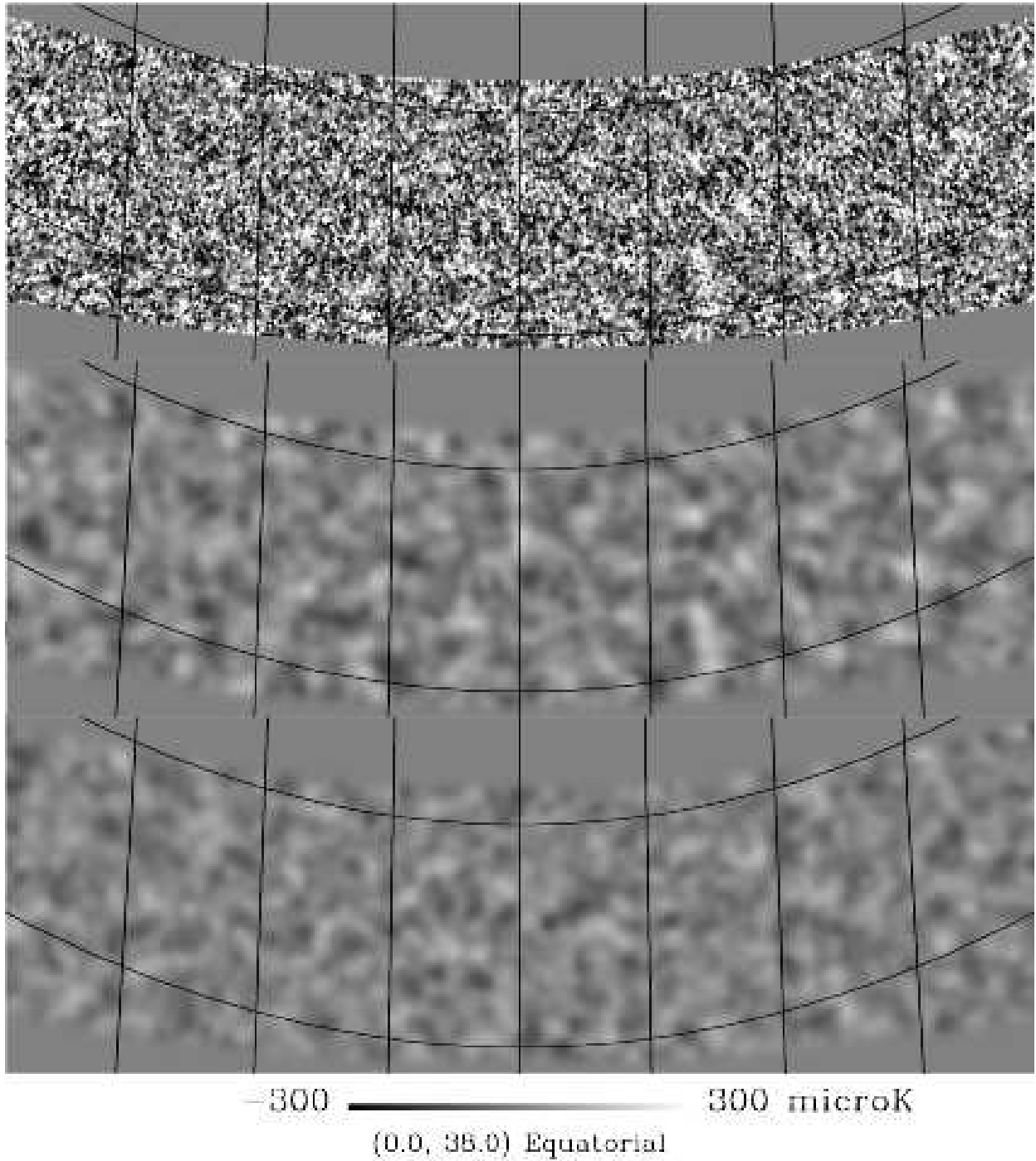}
\caption{ Gnomonic projection of the Q-band map centered at $\alpha$= $0\degr$ and $\delta$= $+38 \degr$.  The top map is the 
full resolution (6.9 arcminute pixels) map. The center map is smoothed with a 30 arcminute FWHM gaussian. The bottom map is 
the first half/second half difference map, also smoothed to 30 arcminutes. The difference map
should have only noise while the average map (center) should show noise+cosmic structure. 
Declination increases up, right ascension increases to the left, the grid is 5 degrees spacing
in right ascension and 10 degrees in declination. \label{fig6}}
\end{figure}
\clearpage

\clearpage
\begin{figure}
\plotone{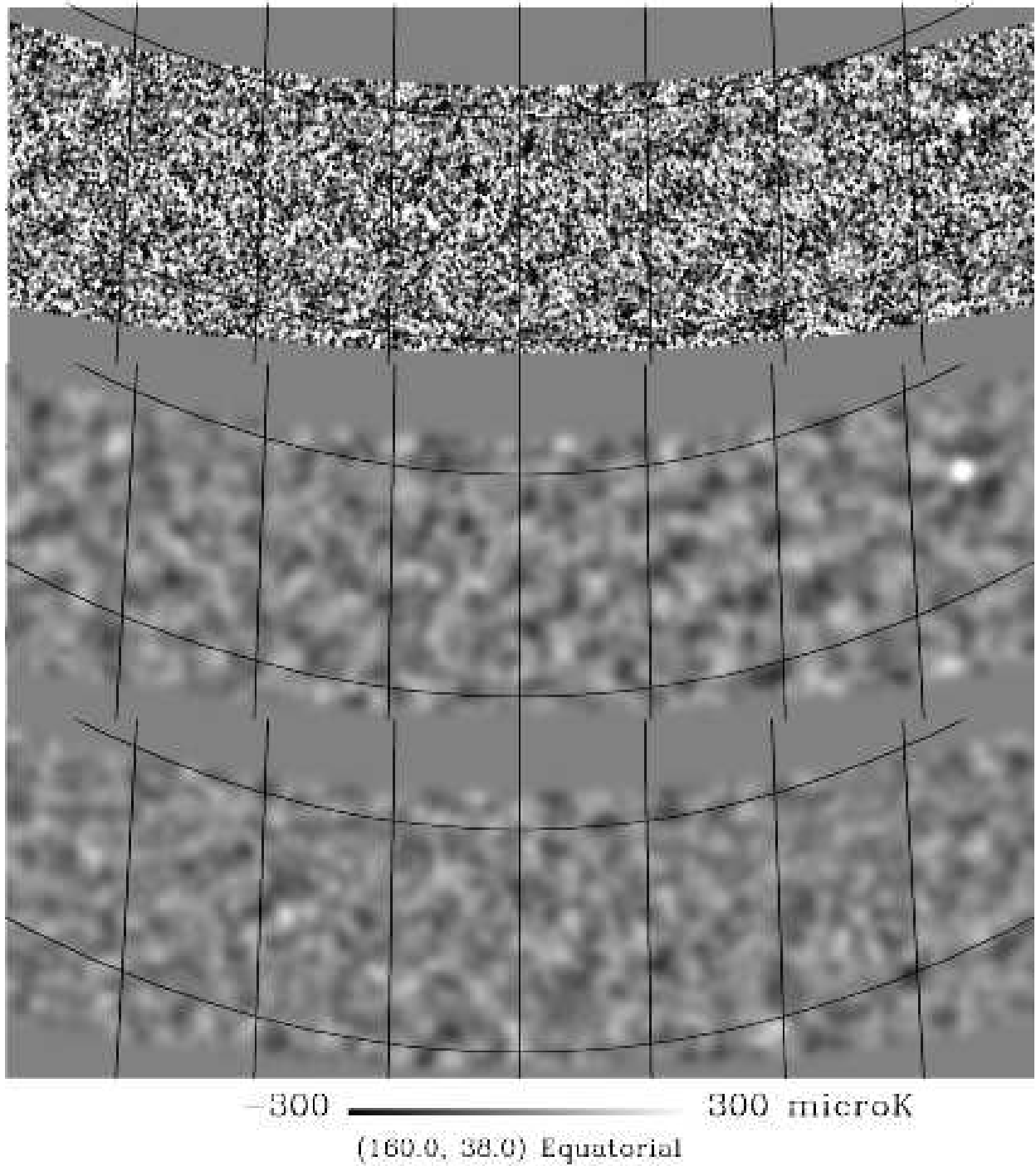}
\caption{ Gnomonic projection of the Q-band map centered at $\alpha$= $160\degr$ and $\delta$= $+38 \degr$.  
This section of the map has an obvious point source at the upper right (removed for power spectrum analysis) \label{fig7}}
\end{figure}
\clearpage

\clearpage
\begin{figure}
\plotone{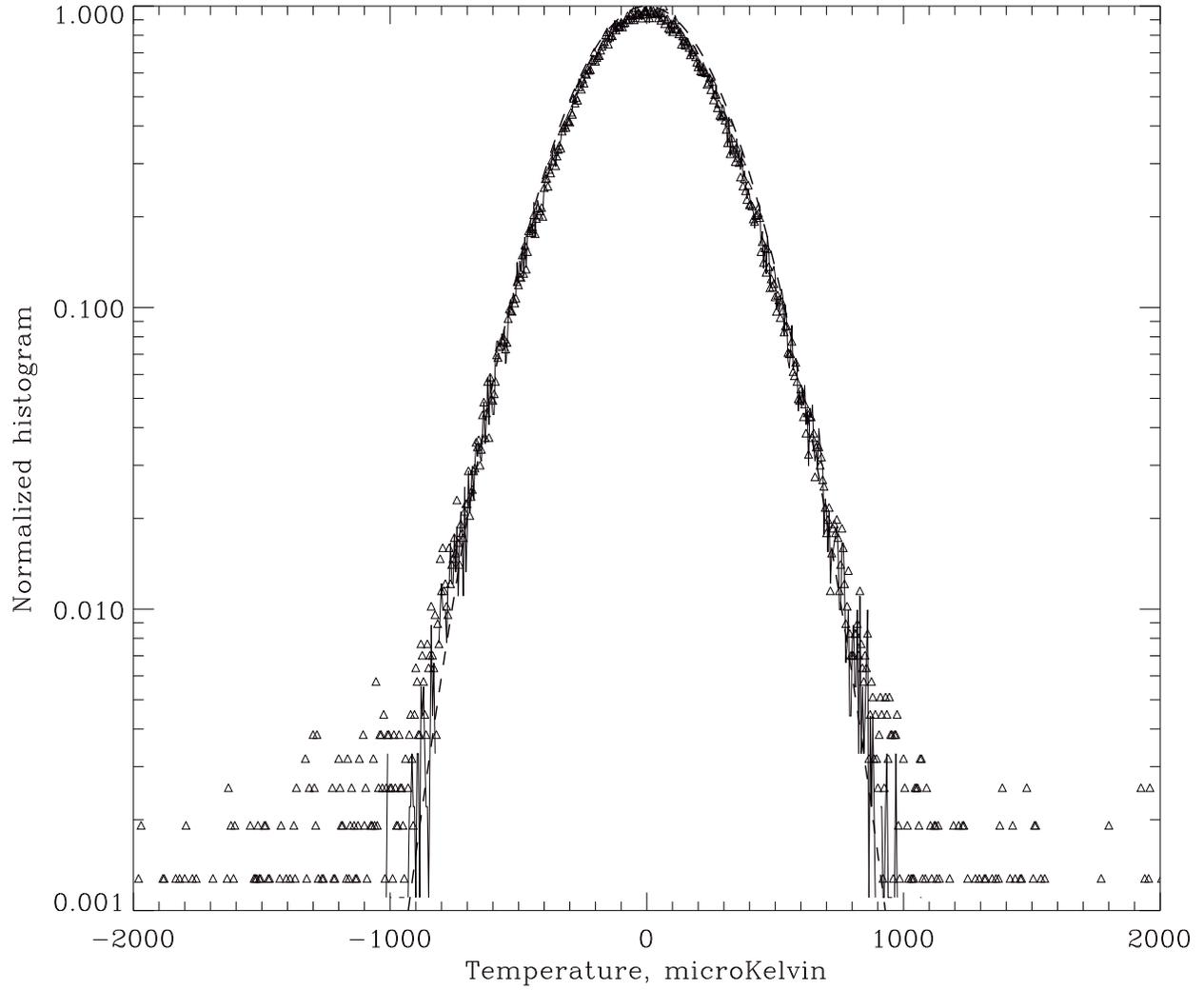}
\caption{ Normalized histogram of the full resolution Q-band map pixels. Symbols are the full map, the solid black
line is the map with galaxy and point sources removed $(\mid b\mid \le 25\degr)$, the dashed line is a 
gaussian with $250 \mu$K $\sigma$. 
\label{fig8}}
\end{figure}
\clearpage 

\clearpage
\begin{figure}
\plotone{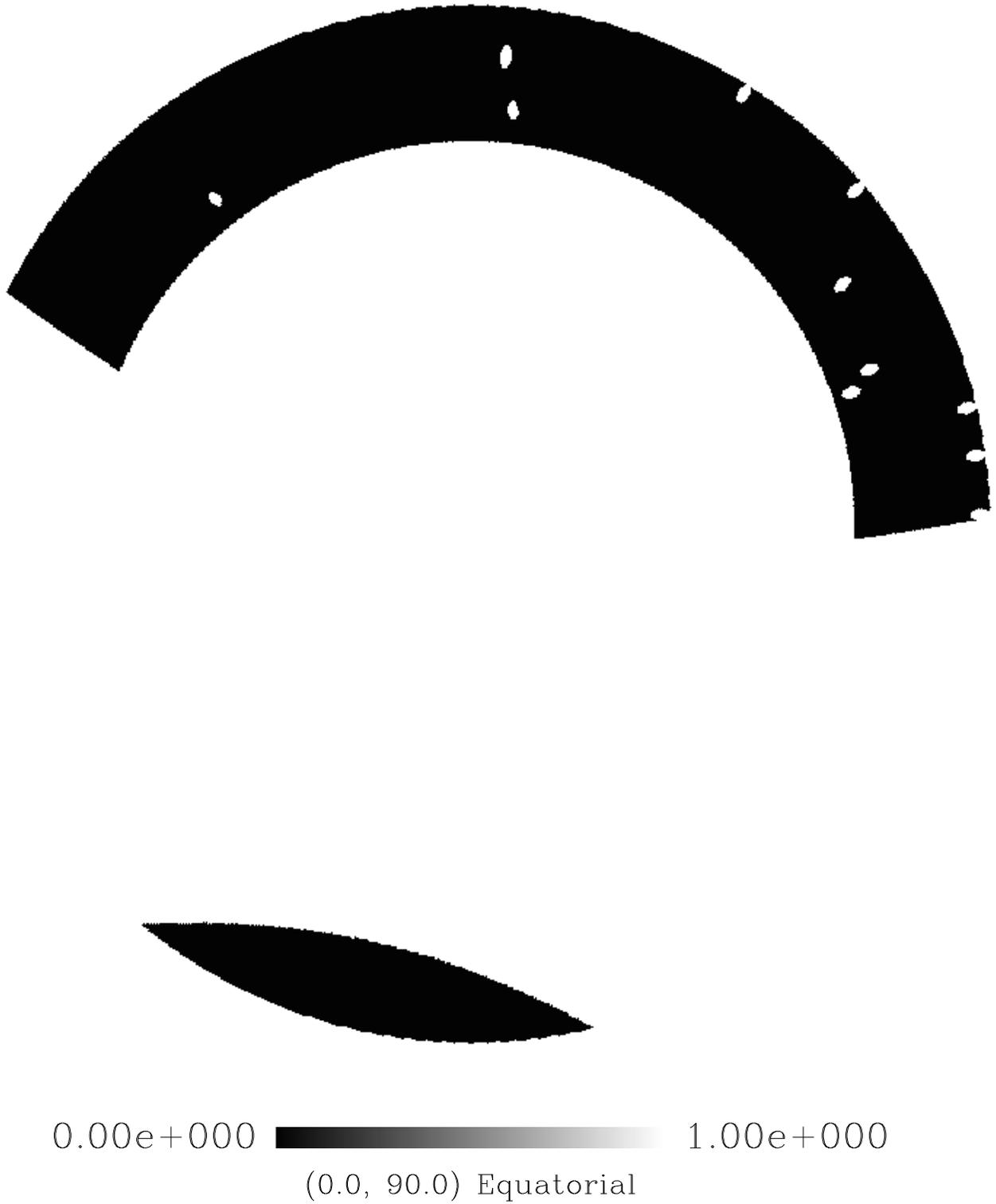}
\caption{ The map sections available for power spectrum analysis after foreground cuts $(\mid b\mid \le 25\degr)$\label{fig9}}
\end{figure}
\clearpage 

\clearpage
\begin{figure}
\plotone{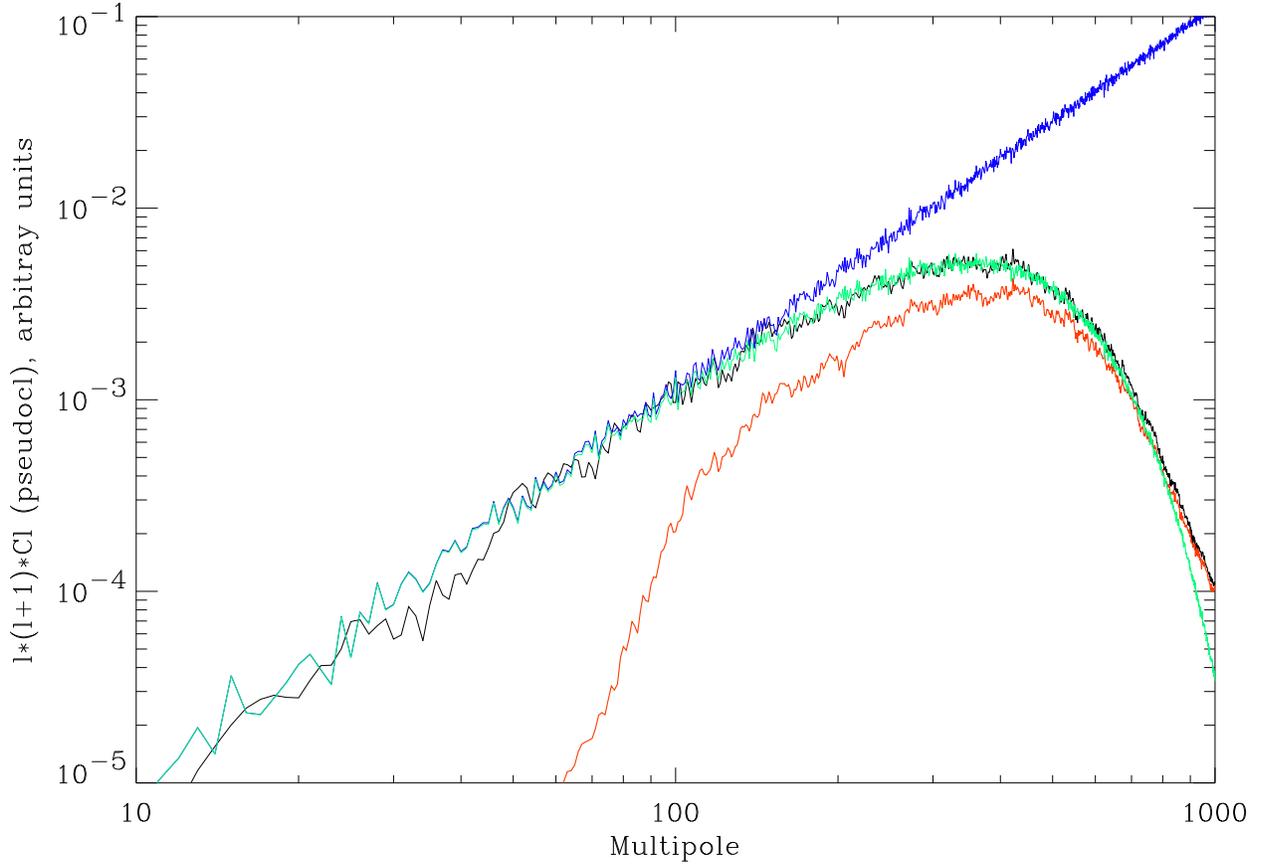}
\caption{ An approximate $\ell$ -space description of the effective filtering produced by our pipeline processing and 
galaxy cuts.  Starting at the top: Power spectrum for the full 'white' simulated sky, scaled by the sky fraction; Spectrum for 
23 arcminute FWHM smoothed simulation, scaled by sky fraction; 23 arcminute smoothed simulation cut by galaxy 
removal and BEAST sky coverage template; Smoothed sky simulation measured through BEAST pipeline analysis 
(no noise added). Note that all spectra were caluclated using the HEALPix 'ANAFAST' routine. Also all except 
the full sky spectrum are 'pseudo-Cl' spectra, calculated from partial sky coverage without any corrections. 
See the text for discussion.
\label{fig10}}
\end{figure}
\clearpage 

\clearpage
\begin{figure}
\plotone{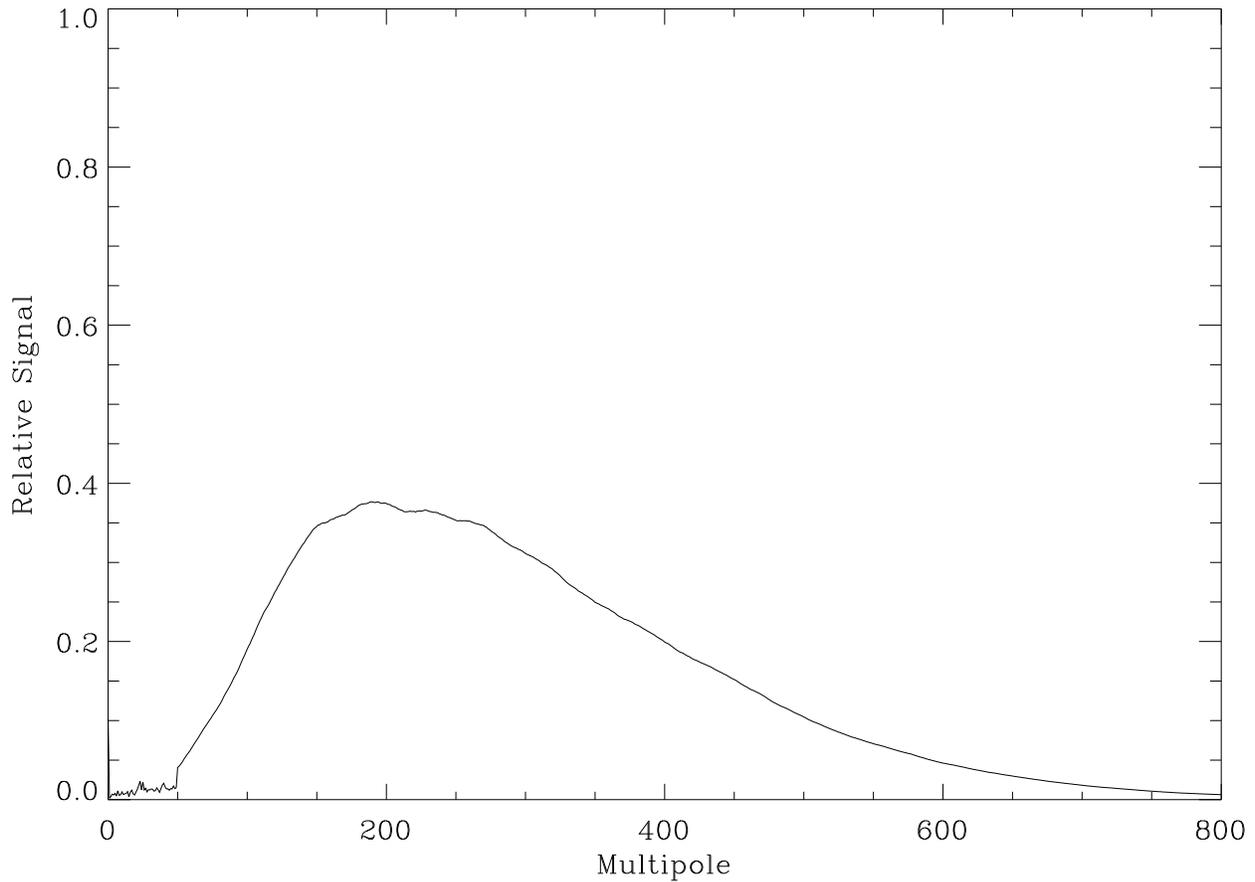}
\caption{ The effective $\ell$ space window function of the experiment and analysis. Plotted is the relative 
sensitivity of the BEAST experiment including beam smoothing, highpass filtering and template removal, as 
compared with a perfect experiment measuring the same region. The comparison is done using the same 
methods as the previous figure. This calculation does not include the effects of the noise, only the
 instrument and analysis filtering.
\label{fig11}}
\end{figure}
\clearpage 

\end{document}